\documentclass[aps,rmp,square]{revtex4}
\usepackage{amsmath,amssymb}
\usepackage{graphicx}
\usepackage{pst-plot}
\usepackage{mathrsfs}
\usepackage{bm}

\begin{document}

\setcitestyle{numbers}

\def\cs#1#2{#1_{\!{}_#2}}
\def\css#1#2#3{#1^{#2}_{\!{}_#3}}
\def\ocite#1{[\citenum{#1}]}
\def\ket#1{|#1\rangle}
\def\bra#1{\langle#1|}
\def\expac#1{\langle#1\rangle}
\def\dbl{\hbox{${1\hskip -2.4pt{\rm l}}$}}
\def\bfh#1{\bf{\hat#1}}

\newenvironment{rcase}
    {\left.\begin{aligned}}
    {\end{aligned}\right\rbrace}

\title{Disproof of Bell's Theorem}

\author{Joy Christian}

\affiliation{Einstein Centre for Local-Realistic Physics, 15 Thackley End, Oxford OX2 6LB, United Kingdom}

\begin{abstract}
We illustrate an explicit counterexample to Bell's theorem by constructing a pair of spin variables in ${\,S^3}$ that exactly reproduces the EPR-Bohm correlation in a manifestly local-realistic manner.\break
\end{abstract}
\maketitle

\parskip -0.25cm

We begin by defining the detections of spin bivectors ${{\bf L}({\bf s},\,\lambda^k)}$ by the detector bivectors ${{\bf D}({\bf a})}$ and ${{\bf D}({\bf b})}$ \{Ref.${\,}$\ocite{IJTP}\}:
\begin{align}
S^3\ni\,{\mathscr A}({\bf a},\,{\lambda^k})\,:=\,\lim_{{\bf s}\,\rightarrow\,{\bf a}}\left\{-\,{\bf D}({\bf a})\,{\bf L}({\bf s},\,\lambda^k)\right\}&=\,                  
\begin{cases}
+\,1\;\;\;\;\;{\rm if} &\lambda^k\,=\,+\,1 \\
-\,1\;\;\;\;\;{\rm if} &\lambda^k\,=\,-\,1
\end{cases} \\
\text{and}\;\;\;\;S^3\ni\,{\mathscr B}({\bf b},\,{\lambda^k})\,:=\,\lim_{{\bf s}\,\rightarrow\,{\bf b}}\left\{+\,{\bf L}({\bf s},\,\lambda^k)\,{\bf D}({\bf b})\right\}&=\,                     
\begin{cases}
-\,1\;\;\;\;\;{\rm if} &\lambda^k\,=\,+\,1 \\
+\,1\;\;\;\;\;{\rm if} &\lambda^k\,=\,-\,1\,,
\end{cases} \label{99-oi}
\end{align}
where the orientation ${\lambda}$ of ${S^3}$ is assumed to be a random variable with 50/50 chance of being ${+1}$ or ${-\,1}$ at the moment of the pair-creation, making the spinning bivector 
${{\bf L}({\bf n},\,\lambda)}$ a random variable {\it relative} to the detector bivector ${{\bf D}({\bf n})}$:
\begin{equation}
{\bf L}({\bf n},\,\lambda)\,=\,\lambda\,{\bf D}({\bf n})\,\,\Longleftrightarrow\,\,{\bf D}({\bf n})\,=\,\lambda\,{\bf L}({\bf n},\,\lambda). \label{OJS}
\end{equation}
The expectation value of the simultaneous outcomes ${{\mathscr A}({\bf a},\,{\lambda^k})=\pm1}$ and ${{\mathscr B}({\bf b},\,{\lambda^k})=\pm1}$ is then worked out as follows: 
\begin{align}
{\cal E}({\bf a},\,{\bf b})\,&=\lim_{\,n\,\gg\,1}\left[\frac{1}{n}\sum_{k\,=\,1}^{n}\,
{\mathscr A}({\bf a},\,{\lambda}^k)\;{\mathscr B}({\bf b},\,{\lambda}^k)\right]\;\;\text{within}\;\;S^3:=\,\text{the set of all unit (left-handed) quaternions} \\
&=\lim_{\,n\,\gg\,1}\left[\frac{1}{n}\sum_{k\,=\,1}^{n}\,\bigg[\lim_{{\bf s}\,\rightarrow\,{\bf a}}\left\{\,-\,{\bf D}({\bf a})\,{\bf L}({\bf s},\,\lambda^k)\right\}\bigg]\left[\lim_{{\bf s}\,\rightarrow\,{\bf b}}\left\{\,+\,{\bf L}({\bf s},\,\lambda^k)\,{\bf D}({\bf b})\right\}\,\right]\right]\;\;\;\text{(conserving total spin = 0)} \\
&=\lim_{\,n\,\gg\,1}\left[\frac{1}{n}\sum_{k\,=\,1}^{n}\,\lim_{\substack{{\bf s}\,\rightarrow\,{\bf a} \\ {\bf s}\,\rightarrow\,{\bf b}}}\left\{\,-\,{\bf D}({\bf a})\,{\bf L}({\bf s},\,\lambda^k)\,\,{\bf L}({\bf s},\,\lambda^k)\,{\bf D}({\bf b})\,\equiv\,{\bf q}({\bf a},\,{\bf b};\,{\bf s},\,\lambda^k)\in S^3\right\}\right] \\
&=\lim_{\,n\,\gg\,1}\left[\frac{1}{n}\sum_{k\,=\,1}^{n}\,\lim_{\substack{{\bf s}\,\rightarrow\,{\bf a} \\ {\bf s}\,\rightarrow\,{\bf b}}}\left\{\,-\,\lambda^k\,{\bf L}({\bf a},\,\lambda^k)\,\,{\bf L}({\bf s},\,\lambda^k)\,{\bf L}({\bf s},\,\lambda^k)\,\,\lambda^k\,{\bf L}({\bf b},\,\lambda^k)\right\}\right] \\
&=\lim_{\,n\,\gg\,1}\left[\frac{1}{n}\sum_{k\,=\,1}^{n}\,\lim_{\substack{{\bf s}\,\rightarrow\,{\bf a} \\ {\bf s}\,\rightarrow\,{\bf b}}}\left\{\,-\,{\bf L}({\bf a},\,\lambda^k)\,\,{\bf L}({\bf s},\,\lambda^k)\,{\bf L}({\bf s},\,\lambda^k)\,\,{\bf L}({\bf b},\,\lambda^k)\right\}\right] \\
&=\lim_{\,n\,\gg\,1}\left[\frac{1}{n}\sum_{k\,=\,1}^{n}\,{\bf L}({\bf a},\,\lambda^k)\,{\bf L}({\bf b},\,\lambda^k)\,\right]\;\;\;\text{\{cf. Appendix B of Ref.${\,}$\ocite{IJTP}\}.}
\label{exppeu}
\end{align}
Here the integrand of (6) is necessarily a unit quaternion ${{\bf q}({\bf a},\,{\bf b};\,{\bf s},\,\lambda^k) \in S^3}$ since ${S^3}$ is closed under multiplication; (7) follows upon using (3); (8) follows upon using ${\lambda^2 = +1}$; and (9) follows from the fact that all unit bivectors such as ${{\bf L}({\bf s},\,\lambda)}$ square to ${-1}$. Using ${I:=\,{\bf e}_x\wedge{\bf e}_y\wedge{\bf e}_z}$ with ${I^2=-1}$, the final sum can now be evaluated by recognizing that the spins in the right and left oriented ${S^3}$ satisfy the following geometrical relations \{cf. Appendix A of Ref.${\,}$\ocite{IJTP}\}:
\begin{align}
{\bf L}({\bf a},\,{\lambda}^k=+1)\;{\bf L}({\bf b},\,{\lambda}^k=+1)\,&=\,(\,+\,I\cdot{\bf a})(\,+\,I\cdot{\bf b}) \\
\text{and}\;\;\;{\bf L}({\bf a},\,{\lambda}^k=-1)\;{\bf L}({\bf b},\,{\lambda}^k=-1)\,&=\,(\,+\,I\cdot{\bf b})(\,+\,I\cdot{\bf a}).
\end{align}
In other words, when ${\lambda^k}$ happens to be equal to ${+1}$, ${{\bf L}({\bf a},\,{\lambda}^k)\;{\bf L}({\bf b},\,{\lambda}^k)=(\,+\,I\cdot{\bf a})(\,+\,I\cdot{\bf b})}$, and when ${\lambda^k}$ happens to be equal to ${-1}$, ${{\bf L}({\bf a},\,{\lambda}^k)\;{\bf L}({\bf b},\,{\lambda}^k)=(\,+\,I\cdot{\bf b})(\,+\,I\cdot{\bf a})}$. Consequently, the above expectation value reduces at once to
\begin{equation}
{\cal E}({\bf a},\,{\bf b})\,=\,\frac{1}{2}(\,+\,I\cdot{\bf a})(\,+\,I\cdot{\bf b})\,+\,\frac{1}{2}(\,+\,I\cdot{\bf b})(\,+\,I\cdot{\bf a})\,
=\,-\,\frac{1}{2}\left\{{\bf a}{\bf b}\,+\,{\bf b}{\bf a}\right\}=\,-\,{\bf a}\cdot{\bf b}\,+\,0\,,\label{stand-nossss}
\end{equation}
because the orientation ${\lambda}$ of ${S^3}$ is a fair coin. Here the last equality follows from the definition of the inner product.

\end{document}